\documentclass[doublespacing]{elsart}

\usepackage{amssymb}
\usepackage{epsfig}
\usepackage{tabularx}
\usepackage[english]{babel}
\usepackage{graphicx}
\usepackage[numbers]{natbib}

\begin{document}
\begin{frontmatter}

\title{Positivity-preserving flux limiters for high-order conservative schemes}


\author[label1]{X. Y. Hu,}  \author[label1]{N. A. Adams} \and \author[label2]{C.-W. Shu}
\address[label1]{Lehrstuhl f\"{u}r Aerodynamik und Str\"{o}mungsmechanik, Technische Universit\"{a}t
M\"{u}nchen,
 85748 Garching, Germany}
\address[label2]{Division of Applied Mathematics, Brown University, Providence, RI 02912, USA}
\begin{abstract}
In this work a simple method
to enforce the positivity-preserving property for general high-order conservative schemes is proposed.
The method keeps the original scheme unchanged and
detects critical numerical fluxes which may lead to negative 
density and pressure,
and then imposes a cut-off flux limiter to satisfy a sufficient 
condition for preserving positivity.
Though an extra time-step size 
condition is required to maintain the formal order of accuracy,
it is less restrictive than those in previous works.
A number of numerical examples suggest that this method, when
applied on an essentially non-oscillatory base scheme, can be 
used to prevent positivity failure when the flow involves 
vacuum or near vacuum and very strong discontinuities.
\end{abstract}
\end{frontmatter}

\section{Introduction} \label{sec:intro}
Compressible flow problems are usually solved by conservative schemes. High-order conservative schemes are suitable for simulating flows with both shock waves and rich flow features (acoustic waves, turbulence) since they are capable of handling flow discontinuities and accurately resolve a broad range of length scales.
One important issue of high-order conservative schemes is that non-physical negative density or pressure (failure of positivity) can lead to an ill-posed system, which may cause
blow-ups of the numerical solution. While for some first-order schemes negative density or pressure can occur when a vacuum or near vacuum is reached, for higher-order conservative schemes positivity failure can also occur due to interpolation errors at or near very strong discontinuities even though the flow physically is far away from vacuum.

It is known that many first order Godunov-type schemes \cite{einfeldt1991godunov, toro2009riemann, gressier1999positivity} have the so called positivity-preserving property and
can maintain positive density and pressure. It has been also proved that some second-order conservative schemes \cite{tao1999gas, hu2004kinetic} 
are positivity-preserving with or without a more restrictive Courant-Friedrichs-Lewy (CFL) condition.
For even higher-order conservative schemes, Perthame and Shu \cite{perthame1996positivity} proved that, given a first-order positivity-preserving scheme, such as Godunov-type schemes, one can always build a higher-order positivity-preserving finite volume scheme under the following constraints: (a) the cell-face values for the numerical flux calculation have positive density and pressure,
(b) additional limits on the interpolation under a more restrictive CFL-like condition.
With a different interpretation of these constraints based on
certain Gauss-Lobatto quadratures, positivity-preserving methods have been successfully developed
for high-order discontinuous Galerkin (DG) methods \cite{zhang2010positivity} and weighted essentially non-oscillatory (WENO) finite volume
and finite difference schemes \cite{zhang2011maximum, zhang2011positivity}.

In this paper, we propose an alternative method to enforce the 
positivity-preserving property with a simple cut-off flux limiter.
The flux limiter first detects critical numerical fluxes which may 
lead to negative density and pressure,
then limits these fluxes to satisfy a sufficient condition for 
preserving positivity.
Unlike the approaches in \cite{zhang2010positivity,zhang2011maximum, zhang2011positivity}, in which positivity-preserving and the maintenance
of high order accuracy are considered simultaneously when designing
the limiter, here we design the cut-off flux limiter to satisfy
positivity only, and then prove {\em a posteriori} the
maintenance of high order accuracy 
under a time step restriction. It appears that, in our numerical 
experiments,
a much less restrictive time-step size condition is sufficient for preserving positivity without destroying overall accuracy.
An advantage of the approach in this paper is that the cut-off limiter
is directly applied to the numerical flux and it
can be applied to arbitrary high-order conservative schemes.

\section{Method}\label{sec:method}
For presentation of the positivity-preserving flux limiters we assume that
the fluid is inviscid and compressible, described by the one-dimensional Euler equations as
\begin{equation}\label{governing-equation}
\frac{\partial \mathbf{U}}{\partial t} + \frac{\partial \mathbf{F}(\mathbf{U})}{\partial x} = 0,
\end{equation}
where $\mathbf{U} = (\rho, m, E)^{T}$, and $\mathbf{F}(\mathbf{U}) = [m, \rho u^2 + p, (E + p)u]^{T}$. This set of equations describes the conservation laws for mass density
$\rho$, momentum density $m \equiv \rho u$ and
total energy density $E=\rho e + \rho u^2/2$, where $e$
is the internal energy per unit mass. To close this set of equations, the ideal-gas equation of state $p = (\gamma -1)\rho e$ with a
constant $\gamma$ is used. Note that the density and pressure have the relations with the conservative variables as
\begin{equation}\label{density-pressure}
\rho(\mathbf{U})  =  \rho,  \quad p(\mathbf{U})  =  (\gamma - 1)\left(E - \frac{1}{2}\frac{m^2}{\rho}\right).
\end{equation} %
It is easy to find that they are locally Lipschitz continuous, i.e.
\begin{eqnarray}
|\rho(\mathbf{U}_2) - \rho(\mathbf{U}_1)| & \leq & L_{\rho}||\mathbf{U}_2 - \mathbf{U}_1||, \label{lipschitz-density} \\ |p(\mathbf{U}_2) - p(\mathbf{U}_1)| & \leq & L_{p}||\mathbf{U}_2 - \mathbf{U}_1||, ~{\rm if} ~\rho(\mathbf{U}_1) > 0,  ~\rho(\mathbf{U}_2)
 > 0, \label{lipschitz-pressure}
\end{eqnarray}
where $L_{\rho}$ and $L_{p}$ are Lipschitz constants. For $1 \geq \theta \geq 0$, $\rho(\mathbf{U})$ and $p(\mathbf{U})$ have the properties
\begin{eqnarray}\label{inequality}
\rho\left[(1 - \theta)\mathbf{U}_1 + \theta \mathbf{U}_2\right] & = & (1 - \theta)\rho(\mathbf{U}_1) + \theta \rho(\mathbf{U}_2), \label{inequality-rho} \\
p\left[(1 - \theta)\mathbf{U}_1 + \theta \mathbf{U}_2\right] & \geq & (1 - \theta)p(\mathbf{U}_1) + \theta p(\mathbf{U}_2),
~{\rm if} ~\rho(\mathbf{U}_1),  \, ~\rho(\mathbf{U}_2) > 0,
\label{inequality-p}
\end{eqnarray} %
where Eq. (\ref{inequality-rho}) is straightforward and
Eq. (\ref{inequality-p}) is implied by the Jensen's inequality
since $p(\mathbf{U})$ is a concave function.
\subsection{Finite-volume and finite-difference conservative schemes}
When Eq. (\ref{governing-equation}) is discretized within the spatial domain such that $x_i =
i\Delta x$, $i=0, ..., N$, where $\Delta x$ is the spatial step,
a general explicit $k$th-order conservative scheme with Euler-forward time integration can be written as
\begin{equation}\label{conservative-scheme}
\mathbf{U}^{n+1}_i = \mathbf{U}^{n}_i + \lambda \left(\hat{\mathbf{F}}_{i-1/2} - \hat{\mathbf{F}}_{i+1/2}\right),
\end{equation}
where the superscript $n$ and $n+1$ represent the old and new time steps, respectively, and $\lambda = \Delta t/\Delta x$, where $\Delta t$ is time-step size.
Note that with the CFL condition
\begin{equation}\label{CFL}
\Delta t = \frac{{\rm CFL} \cdot \Delta x}{(|u| + c)_{\max}},
\end{equation}
where $c=\sqrt{\gamma p /\rho}$ is 
the sound speed and the CFL number $0<{\rm CFL}<1$, one has the relation
\begin{equation}\label{lambda0}
\lambda = \frac{\rm CFL}{(|u| + c)_{\max}}.
\end{equation}
For a finite-volume scheme, $\mathbf{U}^{n}_i$ and $\mathbf{U}^{n+1}_i$
are the cell averaged conservative variables on the cell $i$ defined on the computational cell between $(i-1/2)\Delta x$ and $(i+1/2)\Delta x$,
i.e. $I_i = [x_{i-1/2}, x_{i+1/2}]$,
$\hat{\mathbf{F}}_{i\pm1/2} = \mathbf{F}_{i\pm1/2} + \mathbf{O}(\Delta x^{k+1})$ are the numerical fluxes, which are based on the cell-face values $\mathbf{U}_{i\pm1/2}$
reconstructed from the cell averages $\{ \mathbf{U}_j \}$ and 
$\mathbf{F}_{i\pm1/2} = \mathbf{F}(\mathbf{U}_{i\pm1/2})$.

For a finite-difference scheme, $\mathbf{U}^{n}_i$ and $\mathbf{U}^{n+1}_i$ are the nodal values,
and $(\hat{\mathbf{F}}_{i+1/2} - \hat{\mathbf{F}}_{i-1/2})/\Delta x$ is a $k$th order approximation to $\partial \mathbf{F}(\mathbf{U})/\partial x$ at $x = x_i$.
Assume there exists a function $\mathbf{H}(x)$ depending on $\Delta x$ such that
\begin{equation}\label{reconstruction-pair}
\mathbf{F}\left[\mathbf{U}(x)\right] = \frac{1}{\Delta x}\int^{x + \Delta x /2}_{x - \Delta x /2} \mathbf{H}(\xi) d \xi,
\end{equation}
then the same reconstruction
procedure in a finite-volume scheme can be used to obtain
the numerical fluxes $\hat{\mathbf{F}}_{i\pm1/2} = \mathbf{H}_{i\pm1/2} + O(\Delta x^{k+1})$ based on the cell-face values of $\mathbf{H}(x)$
reconstructed from its cell-average values
$\mathbf{F}\left[\mathbf{U}_j\right] =  \int^{x_j 
+ \Delta x /2}_{x_j - \Delta x /2} \mathbf{H}(\xi) d \xi /\Delta x $.
We refer to \cite{SO2} for the discussion of this formulation of
conservative finite difference schemes.
\subsection{Positivity preserving cut-off flux limiter}
The positivity-preserving property for the 
scheme Eq. (\ref{conservative-scheme})
refers to the property that the density and pressure are
positive for
$\mathbf{U}^{n+1}_i$ when  $\mathbf{U}^{n}_i$ has positive 
density and pressure.
Since Eq. (\ref{conservative-scheme}) can be rewritten as a 
convex combination
\begin{eqnarray}\label{rewritten-scheme}
\mathbf{U}^{n+1}_i & = & \frac{1}{2}\left(\mathbf{U}^{n}_i + 2\lambda \hat{\mathbf{F}}_{i-1/2} \right)  + \frac{1}{2}\left(\mathbf{U}^{n}_i - 2\lambda \hat{\mathbf{F}}_{i+1/2}\right) \nonumber \\
 & = & \frac{1}{2}\mathbf{U}^{-}_i + \frac{1}{2}\mathbf{U}^{+}_i, \end{eqnarray}
a sufficient condition for preserving positivity is that $\mathbf{U}^{\pm}_i$ have positive density and pressure, i.e. $g(\mathbf{U}^{\pm}_i)>0$, where $g$ represents $\rho$ and $p$.
Since the first-order Lax-Friedrichs flux %
\begin{equation}\label{Lax-Friedrichs}
\hat{\mathbf{F}}^{LF}_{i+1/2} = \frac{1}{2}\left[\mathbf{F}_{i} + \mathbf{F}_{i+1} + (|u| + c)_{max}(\mathbf{U}^{n}_i - \mathbf{U}^{n}_{i+1})\right]
\end{equation}
has the property $g(\mathbf{U}^{LF,\pm}_i) = g(\mathbf{U}^{n}_i \mp 2\lambda \hat{\mathbf{F}}^{LF}_{i \pm 1/2})>0$, under an additional CFL condition
\begin{equation}\label{CFL-condition-1}
{\rm CFL} \le \frac{1}{2}
\end{equation}
(see \cite{zhang2010positivity}), a straightforward way to ensure 
positivity is to limit the magnitude of $\hat{\mathbf{F}}_{i+1/2}$
by utilizing the properties in 
Eqs. (\ref{inequality-rho}) and (\ref{inequality-p}).
The positive density is first enforced by:
\begin{description}
        \item[Cut-off flux limiter for positive density]
        \item[1.] For all $i$: initialize $\theta^{+}_{i+1/2} = 1$, $\theta^{-}_{i+1/2} = 1$.
        \item[2.] If $ \rho(\mathbf{U}^{+}_i)<\epsilon_{\rho}$ , solve $\theta^{+}_{i+1/2}$
        from $(1 - \theta^{+}_{i+1/2})\rho(\mathbf{U}^{LF,+}_i)  +  \theta^{+}_{i+1/2} \rho(\mathbf{U}^{+}_i) = \epsilon_{\rho}$.
        \item[3.] If $ \rho(\mathbf{U}^{-}_{i+1})<\epsilon_{\rho}$, solve $\theta^{-}_{i+1/2}$
        from $(1 - \theta^{-}_{i+1/2})\rho(\mathbf{U}^{LF,-}_{i+1}) + \theta^{-}_{i+1/2} \rho(\mathbf{U}^{-}_{i+1}) = \epsilon_{\rho}$.
  \item[4.] Set $\theta_{\rho,i+1/2} = \min(\theta^{+}_{i+1/2}, \theta^{-}_{i+1/2})$, $\hat{\mathbf{F}}^{*}_{i+1/2} = (1-\theta_{\rho,i+1/2})\hat{\mathbf{F}}^{LF}_{i+1/2} + \theta_{\rho,i+1/2}  \hat{\mathbf{F}}_{i+1/2}$.
\end{description}
Here, $\epsilon_{\rho} = \min\left\{10^{-13}, \rho^{0}_{min}\right\}$, where $\rho^{0}_{min}$ is the minimum density in the initial condition, $\hat{\mathbf{F}}^{*}_{i+1/2}$ is the limited flux, $0 \leq \theta^{\pm}_{i+1/2} \leq 1$ are the limiting factors corresponding to the two neighboring cells, which share the same flux $\hat{\mathbf{F}}_{i+1/2}$. After applying 
this flux limiter, Eq. (\ref{rewritten-scheme}) becomes
\begin{eqnarray}\label{rewritten-scheme-1}
\mathbf{U}^{n+1}_i & = & \frac{1}{2}\left(\mathbf{U}^{n}_i + 2\lambda \hat{\mathbf{F}}^{*}_{i-1/2} \right)  + \frac{1}{2}\left(\mathbf{U}^{n}_i - 2\lambda \hat{\mathbf{F}}^{*}_{i+1/2}\right) \nonumber \\
 & = & \frac{1}{2}\mathbf{U}^{*,-}_i + \frac{1}{2}\mathbf{U}^{*,+}_i. \end{eqnarray}
Clearly, by Eq. (\ref{inequality-rho}),
both $\mathbf{U}^{*,-}_i$ and $\mathbf{U}^{*,+}_i$ have positive density,
so does $\mathbf{U}^{n+1}_i$.
The positive pressure is further enforced by:
\begin{description}
        \item[Cut-off flux limiter for positive pressure]
        \item[1.] For all $i$: initialize $\theta^{+}_{i+1/2} = 1$, $\theta^{-}_{i+1/2} = 1$.
        \item[2.] If $ p(\mathbf{U}^{*,+}_i)<\epsilon_{p}$ , solve $\theta^{+}_{i+1/2}$
        from $(1 - \theta^{+}_{i+1/2})p(\mathbf{U}^{LF,+}_i)  +  \theta^{+}_{i+1/2} p(\mathbf{U}^{*,+}_i) = \epsilon_{p}$.
        \item[3.] If $ p(\mathbf{U}^{*,-}_{i+1})<\epsilon_{p}$, solve $\theta^{-}_{i+1/2}$
        from $(1 - \theta^{-}_{i+1/2})p(\mathbf{U}^{LF,-}_{i+1}) + \theta^{-}_{i+1/2} p(\mathbf{U}^{*,-}_{i+1}) = \epsilon_{p}$.
  \item[4.] Set $\theta_{p,i+1/2} = \min(\theta^{+}_{i+1/2}, \theta^{-}_{i+1/2})$, $\hat{\mathbf{F}}^{**}_{i+1/2} = (1-\theta_{p,i+1/2})\hat{\mathbf{F}}^{LF}_{i+1/2} + \theta_{p,i+1/2}  \hat{\mathbf{F}}^{*}_{i+1/2}$.
\end{description}
Again, $\epsilon_{p} = \min\left\{10^{-13}, p^{0}_{min}\right\}$, where $p^{0}_{min}$ is the minimum pressure in the initial condition, and $\hat{\mathbf{F}}^{**}_{i+1/2}$ is the further limited flux.
After applying 
this flux limiter, Eq. (\ref{rewritten-scheme-1}) becomes
\begin{eqnarray}\label{conservative-scheme-limited}
\mathbf{U}^{n+1}_i & = &  
\frac 12 \left(\mathbf{U}^{n}_i + 2\lambda 
\hat{\mathbf{F}}^{**}_{i-1/2} \right)  + \frac 12 \left(\mathbf{U}^{n}_i - 
2\lambda \hat{\mathbf{F}}^{**}_{i+1/2}\right) \nonumber \\
 & = & \frac 12 \left( \mathbf{U}^{**,-}_i + \mathbf{U}^{**,+}_i
\right).
\end{eqnarray}
Clearly, by Eqs. (\ref{inequality-rho}) and (\ref{inequality-p}),
both $\mathbf{U}^{**,-}_i$ and $\mathbf{U}^{**,+}_i$ have positive density and pressure,
so does $\mathbf{U}^{n+1}_i$.
Note that these limiters can be applied at each sub-stage of a TVD Runge-Kutta \cite{shu1988efficient} method,
which is a convex combination of Euler-forward time steps. 
\subsection{Consistency and accuracy}
Now we address two important issues for the cut-off flux limiter.
First, the limited flux is a consistent flux since it is the convex combination of two consistent fluxes, i.e. the first-order Lax-Friedrichs flux $\mathbf{U}^{LF}_{i+1/2}$ and the original high-order numerical flux $\hat{\mathbf{F}}^{o}_{i+1/2}$, which represents $\hat{\mathbf{F}}_{i+1/2}$ and $\hat{\mathbf{F}}^{*}_{i+1/2}$. Second, when the limiter is active, the difference between the original flux $\hat{\mathbf{F}}^{o}_{i+1/2}$ and the limited flux $\hat{\mathbf{F}}^{lim}_{i+1/2}$ representing $\hat{\mathbf{F}}^{*}_{i+1/2}$ and $\hat{\mathbf{F}}^{**}_{i+1/2}$ is
\begin{equation}\label{correction}
||\hat{\mathbf{F}}^{lim}_{i+1/2} - \hat{\mathbf{F}}^{o}_{i+1/2}|| = (1 - \theta_{g,i+1/2})||\hat{\mathbf{F}}^{o}_{i+1/2} - \hat{\mathbf{F}}^{LF}_{i+1/2}||.
\end{equation}
We only need to consider accuracy maintenance when 
$\theta_{g,i+1/2} <1$, for otherwise the limiter does not take
any effect.  Without loss of generality we may assume
$\theta_{g,i+1/2} = \theta^+_{g,i+1/2}$.  In this situation we have 
$g(\mathbf{U}^{o,+}_i) < \epsilon_{g}$, 
in which $\mathbf{U}^{o,+}_i$ represents $\mathbf{U}^{+}_i$ 
and $\mathbf{U}^{*,+}_i$, and $\epsilon_{g}$ is 
negligibly small, and
\begin{equation}\label{accuracy-condition}
1 - \theta_{g,i+1/2} = \frac{\epsilon_{g} - g(\mathbf{U}^{o,+}_i)}
{g(\mathbf{U}^{LF,+}_i) - g(\mathbf{U}^{o,+}_i)} 
\approx \frac{ - g(\mathbf{U}^{o,+}_i)}
{g(\mathbf{U}^{LF,+}_i) - g(\mathbf{U}^{o,+}_i)} 
\leq
\frac{|g(\mathbf{U}^{o,+}_i)|}
{g(\mathbf{U}^{LF,+}_i)} .
\end{equation}
Since $\hat{\mathbf{F}}^{o}_{i+1/2}$ and 
$\hat{\mathbf{F}}^{LF}_{i+1/2}$ are both bounded in smooth regions,
it is sufficient to show that the accuracy is not destroyed if 
the limiting factor satisfies 
\begin{equation}
\label{accuracy-condition2}
1 - \theta_{g,i+1/2} = O(\Delta x^{k+1}),
\end{equation}
a sufficient condition for which would be
$|g(\mathbf{U}^{o,+}_i)| = O(\Delta x^{k+1})$
and $g(\mathbf{U}^{LF,+}_i)$ is bounded away from zero.

Similar to Zhang and Shu \cite{zhang2010positivity}, 
we assume the exact solution $\mathbf{U}(x)$ is smooth 
and $g(\widetilde{\mathbf{U}}_i) \geq M$, where $\widetilde{\mathbf{U}}_i$ 
is either the cell-average (for the finite-volume scheme) 
or the nodal value (for the finite-different scheme) 
of the exact solution $\mathbf{U}(x)$ and $M>0$ is a constant. 
Since $g(\mathbf{U}_i)$ is obtained from a $k$th order
approximation, one has $g(\mathbf{U}_i)\geq M - O(\Delta x^{k+1}) > M/2$ 
if $\Delta x$ is sufficiently small, therefore 
\begin{eqnarray}\label{bounding}
g(\mathbf{U}^{LF,+}_i) & = & g\left[(1-\hat{w})\mathbf{U}_i + \hat{w}\left(\mathbf{U}_i- \frac{2\lambda}{\hat{w}} \hat{\mathbf{F}}^{LF}_{i+1/2}\right)\right] \nonumber \\
& \geq & (1-\hat{w})g(\mathbf{U}_i) + \hat{w}g\left(\mathbf{U}_i- \frac{2\lambda}{\hat{w}} \hat{\mathbf{F}}^{LF}_{i+1/2}\right) \\
& \geq & \frac{(1-\hat{w})}{2} M >0, \nonumber
\end{eqnarray}
where $1>\hat{w}>0$ is a constant, under an extra CFL condition 
\begin{equation}\label{CFL-condition-2}
{\rm CFL} \le \frac{\hat{w}}{2}.
\end{equation}
Furthermore, one has 
\begin{eqnarray}\label{rewritten-positivity}
\mathbf{U}^{o,+}_i & = & \mathbf{U}^{n}_i - 
2\lambda \hat{\mathbf{F}}^{o}_{i+1/2}\nonumber \\
 & = & \mathbf{U}^{LF,+}_i  +   2\lambda 
\left(\hat{\mathbf{F}}^{LF}_{i+1/2} - 
\hat{\mathbf{F}}^{o}_{i+1/2} \right) \nonumber \\  
 & = & \mathbf{U}^{LF,+}_i  +   2\lambda 
\left(\hat{\mathbf{F}}^{LF}_{i+1/2} - 
\widetilde{\mathbf{F}}_{i+1/2}\right)  
+  \mathbf{O}(\Delta x^{k+1}), \end{eqnarray}
where $\widetilde{\mathbf{F}}_{i+1/2} = \mathbf{F}_{i+1/2}$ 
for the finite-volume scheme, 
and $\widetilde{\mathbf{F}}_{i+1/2} = \mathbf{H}_{i+1/2}$ for 
the finite-difference scheme.
Let $\mathbf{U}^{s}_i = \mathbf{U}^{LF,+}_i  +   2\lambda 
\left(\hat{\mathbf{F}}^{LF}_{i+1/2} - 
\widetilde{\mathbf{F}}_{i+1/2}\right)$, 
and with Eqs. (\ref{lipschitz-density}) and (\ref{lipschitz-pressure}),
one has 
\begin{equation}\label{rewritten-positivity-1}
|g(\mathbf{U}^{s}_i) - g(\mathbf{U}^{o,+}_i)|  
\leq L_{g}||\mathbf{U}^{o,+}_i - \mathbf{U}^{s}_i|| =  O(\Delta x^{k+1}),
\end{equation}
where $L_{g}$ is the Lipschitz constant. 
Note that the first term of $\mathbf{U}^{s}_i$ 
has positive density and pressure.
For the second term, 
since the first-order Lax-Friedrichs flux 
$\hat{\mathbf{F}}^{LF}_{i+1/2}$ is 
a first order approximation to the exact flux
$\widetilde{\mathbf{F}}_{i+1/2}$,
that is $||\hat{\mathbf{F}}^{LF}_{i+1/2} - 
\widetilde{\mathbf{F}}_{i+1/2}|| = O(\Delta x) $.
With bounded $g(\mathbf{U}^{LF,+}_i)$ from Eq. (\ref{bounding}), 
one has 
$$
\rho(\mathbf{U}^{s}_i) \geq \frac{(1-\hat{w})}{2} M - O(\Delta x)
 \geq \frac{(1-\hat{w})}{4} M >0 
$$
for sufficiently small $\Delta x$, according to 
Eq. (\ref{lipschitz-density}), 
and furthermore $p(\mathbf{U}^{s}_i) > \epsilon_{p}$ according 
to Eq. (\ref{lipschitz-pressure}).
Since $g(\mathbf{U}^{s}_i) > \epsilon_{g}$ and $g(\mathbf{U}^{o,+}_i) < \epsilon_{g}$, 
i.e. $\rho(\mathbf{U}^{+}_i) < \epsilon_{\rho}$ while enforcing positive density and $p(\mathbf{U}^{*,+}_i) < \epsilon_{p}$ 
but $\rho(\mathbf{U}^{*,+}_i) > \epsilon_{\rho}$ while enforcing positive pressure, 
Eq. (\ref{rewritten-positivity-1}) leads to
$|g(\mathbf{U}^{o,+}_i)|= O(\Delta x^{k+1})$. 
Hence, we have proved that the cut-off flux limiter preserves high-order accuracy.

Note that, for given values of $M$ and grid size, Eqs. (\ref{bounding}) 
and (\ref{rewritten-positivity}) suggest that the errors introduced by 
the the cut-off flux limiter decrease with the time-step sizes.
Also note that, the condition Eq. (\ref{CFL-condition-2}) is less restrictive than 
the time-step size conditions in Refs. \cite{zhang2010positivity,zhang2011maximum, zhang2011positivity},
and is desirable for higher computational efficiency.
\subsection{Assessment of accuracy}
As a simple way to test the accuracy of the present flux limiters,
we consider the one-dimensional linear advection equation
\begin{equation}
\frac{\partial u}{\partial t} + \frac{\partial u}{\partial x} 
= 0 \label{1d-linear-advection}
\end{equation}
with initial condition $u(x)>0$. 
Applying the cut-off flux limiter to preserve positivity 
results in the limiter (denoted as HAS)
\begin{eqnarray}
f^{*}_{i+1/2} & = & \theta(u_{i+1/2}^- - u^{n}_i) + 
u^{n}_i, \quad \theta = \min \left\{\frac{u^{n}_i}{u^{n}_i - u_{\min}}, 
1\right\}, \nonumber\\ u_{\min} & =  &\min \left\{u^{n}_i - 
2\lambda u_{i+1/2}^-, u^{n}_{i+1} + 2\lambda u_{i+1/2}^-, 
10^{-13}\right\}. \label{1d-linear-advection-limiter-1}
\end{eqnarray}
Here $u_{i+1/2}^-$ is the approximated upwind flux at 
the cell face ${i+1/2}$. Note that only one of 
$u^{n}_i - 2\lambda u_{i+1/2}^-$ or
$u^{n}_{i+1} + 2\lambda u_{i+1/2}^-$ being negative will 
activate the limiter.
The limiter of Zhang and Shu \cite{zhang2011maximum} 
(denoted as ZS) for Eq. (\ref{1d-linear-advection}) can be written as
\begin{eqnarray}
f^{*}_{i+1/2} & = & \theta(u^{-}_{i+1/2} - u^{n}_i) + u^{n}_i, 
\quad \theta = \min \left\{\frac{u^{n}_i}{u^{n}_i 
- u_{\min}}, 1\right\}, \nonumber\\ u_{\min} & 
=  & \min \left\{\frac{u^{n}_i - \hat{w}_1 (u^{+}_{i-1/2} 
+ u^{-}_{i+1/2})}{1-2\hat{w}_1}, u^{+}_{i-1/2}, u^{-}_{i+1/2}, 
10^{-13}\right\}, \label{1d-linear-advection-limiter-shu}
\end{eqnarray}
where $u^{+}_{i-1/2}$ and $u^{-}_{i+1/2}$ are the high-order 
approximations of the cell-face values at $u(x_{i-1/2})$ and 
$u(x_{i+1/2})$ within cell $i$. For Eq. (\ref{1d-linear-advection}), 
one has $u^{+}_{i-1/2} = u_{i+1/2}$. Comparing 
$u_{\min}$ in Eqs. (\ref{1d-linear-advection-limiter-1}) 
and (\ref{1d-linear-advection-limiter-shu}),
it can be observed that the HAS limiter does not directly 
constrain the cell-face values to be non-negative.

To further illustrate the accuracy of the HAS limiter
and its relation to the ZS limiter, we compute the advection 
of a function $u = 1 + 10^{-6} + \cos(2\pi x)$
in domain [0, 1] with a fifth-order conservative finite difference WENO-5 
scheme \cite{jiang1996cient} with third-order TVD Runge-Kutta time 
integration \cite{shu1988efficient}.
A periodic boundary condition is applied at $x=0$ and
$x=1$. The final time is $t=1$, which corresponds to one period. %
\begin{figure}[p]
\begin{center}
\includegraphics[width=1.2\textwidth]{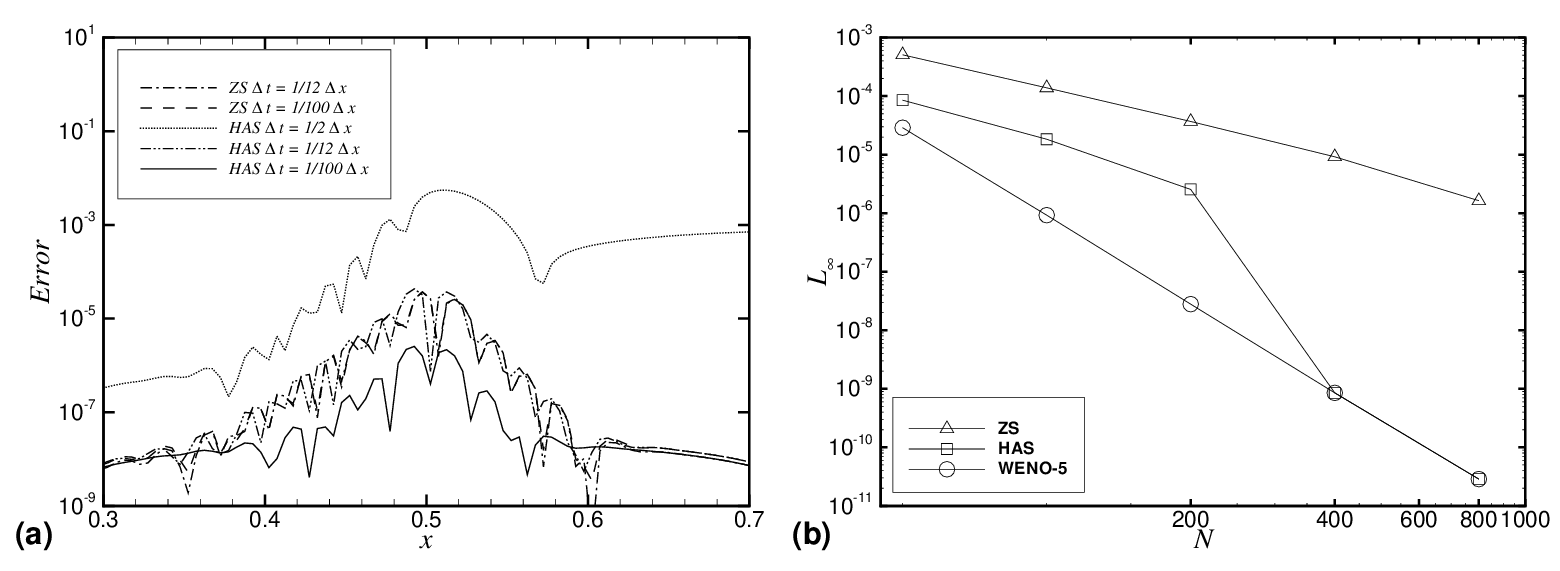}
\caption{Linear advection problem at $t=1$:
(a) Error distribution vs. time-step sizes on 200 grid points; 
(b) Evolution of the $L_\infty$ error with decreasing grid size..} 
\label{1d-advection-weno5}
\end{center}
\end{figure}
This problem is computed on different grids with $N = 50$, 100, 200, 
400 and 800 grid points. Figure \ref{1d-advection-weno5}a shows 
the error distributions for the results on 200 grid points. It can 
be observed that if the maximum admissible CFL number of 0.5 is 
used, the HAS limiter produces larger errors than the ZS limiter.
However,
the HAS limiter is already as accurate as the ZS limiter 
when a smaller CFL number of $1/12$, 
which corresponds to the maximum admissible value for the latter, is used.
If the time-step size is decreased further, errors produced by 
the ZS limiter do not change considerably,
whereas the errors produced by the HAS limiter decrease further.
This behavior is also shown in Fig. \ref{1d-advection-weno5}b for 
the evolution of the $L_\infty$ error with decreasing grid size. 
Here, the time-step size $\Delta t = 0.5\Delta x^{5/3}$ is used 
to keep the spatial errors dominant.
Note that Fig. \ref{1d-advection-weno5}b clearly shows that
the theoretical order of accuracy is achieved.
\subsection{Extension to multiple dimensions}
To present the extension of the positivity-preserving flux 
limiters to multiple dimensions we consider the two-dimensional 
Euler equation
\begin{equation}\label{governing-equation-2d}
\frac{\partial \mathbf{U}}{\partial t} + \frac{\partial 
\mathbf{F}(\mathbf{U})}{\partial x} 
+ \frac{\partial \mathbf{G}(\mathbf{U})}{\partial y} = 0.
\end{equation}
where $\mathbf{U} = (\rho, \rho u, \rho v, E)^{T}$, 
$\mathbf{F}(\mathbf{U}) = [\rho u, \rho u^2 + p, \rho u v, 
(E + p)u]^{T}$ and $\mathbf{G}(\mathbf{U}) = 
[\rho v, \rho v u, \rho v^2 + p, (E + p)v]^{T}$. Compared to 
the one-dimensional equation Eq. (\ref{governing-equation}), 
the momentum density is $\rho\mathbf{v} = (\rho u, \rho v)$, 
where $u$ and $v$ are velocities in the $x$ and $y$ directions, 
respectively, and the total energy density is $E=\rho e +  
\rho \mathbf{v}^2/2$.
As an extension of Eq. (\ref{rewritten-scheme}), the 
conservative scheme for Eq. (\ref{governing-equation-2d}) 
can be rewritten as a convex combination
\begin{eqnarray}\label{rewritten-scheme-multi}
\mathbf{U}^{n+1}_{i,j} & = & \frac{\alpha_x}{2}\left(\mathbf{U}^{n}_{i,j} 
+ 2\lambda_x \hat{\mathbf{F}}_{i-1/2, j}\right) 
+ \frac{\alpha_x}{2}\left(\mathbf{U}^{n}_{i,j}  - 
2\lambda_x\hat{\mathbf{\mathbf{F}}}_{i+1/2, j}\right) \nonumber \\
& + & \frac{\alpha_y}{2}\left(\mathbf{U}^{n}_{i,j} 
+ 2\lambda_y \hat{\mathbf{G}}_{i, j-1/2}\right)  
+ \frac{\alpha_y}{2}\left(\mathbf{U}^{n}_{i,j}  - 
2\lambda_y \hat{\mathbf{G}}_{i, j+1/2}\right),
\end{eqnarray}
where $\lambda_x = \Delta t/\Delta x \alpha_x$ and 
$\lambda_y = \Delta t/\Delta y \alpha_y$, $\alpha_x + \alpha_y = 1$, 
with $\alpha_x>0$ and $\alpha_y>0$ being partitions of
the contribution in the $x$ and $y$ directions.
A simple way to obtain this partition is to set 
$\alpha_x = \alpha_y = 1/2$ as in Zhang and Shu 
\cite{zhang2010positivity, zhang2011positivity}.
Another straightforward way to determine $\alpha_x$ and 
$\alpha_y = 1- \alpha_x$ is
\begin{equation}\label{partition}
\alpha_x = \frac{\tau_x}{\tau_x + \tau_y},
\quad \tau_x = \frac{(|u| + c)_{\max}}{\Delta x}, \quad 
\tau_y = \frac{(|v| + c)_{\max}}{\Delta y}.
\end{equation}
Note that, since the time-step size for integrating Eq. 
(\ref{rewritten-scheme-multi}) is given by
\begin{equation}\label{CFL-2d}
\Delta t = \frac{\rm CFL}{\tau_x + \tau_y},
\end{equation}
one has the relation
\begin{equation}\label{lambda-2d}
\lambda_x = \frac{\rm CFL}{(|u| + c)_{\max}} \quad {\rm and} 
\quad \lambda_y = \frac{\rm CFL}{(|v| + c)_{\max}},
\end{equation}
which gives an extended form from Eq. (\ref{lambda0}).
Also note that, since the components in Eq. 
(\ref{rewritten-scheme-multi}) and Eq. (\ref{rewritten-scheme}) 
have the same form, it is straightforward to implement the 
positivity-preserving flux limiters in a dimension-by-dimension fashion.
\section{Test cases}
In the following, we illustrate that a number of typical numerical 
test cases, where the original high-order conservative schemes fail,
can be simulated by using the proposed positivity-preserving flux 
limiters. For the first type of cases involving vacuum or near 
vacuum, the flux limiters are combined with the finite
difference WENO-5 scheme \cite{jiang1996cient}, which is a 
shock-capturing scheme with fifth-order accuracy for smooth 
solutions. For the second type of cases involving very strong 
discontinuities,
the flux limiters are combined with the WENO-CU6-M1 scheme 
\cite{hu2011scale},
which can be used for implicit large eddy simulation (LES) of 
turbulent flow and has sixth-order accuracy for smooth solutions.
For both variants of the WENO schemes the Roe approximation is 
used for the characteristic decomposition at the cell faces, 
the Lax-Friedrichs formulation is used for the numerical fluxes, and
the third-order TVD Runge-Kutta scheme is used for time 
integration \cite{shu1988efficient}.
If not mentioned otherwise, the computations are carried out 
with a CFL number of 0.5.
\subsection{One-dimensional problems involving vacuum or near vacuum}
Here we show that the proposed method passes two one-dimensional 
test problems involving vacuum or near vacuum: the double 
rarefaction problem  \cite{hu2004kinetic},
where a vacuum occurs, and the planar Sedov blast-wave problem 
\cite{sedov1959similarity, zhang2011positivity}, where a 
point-blast wave propagates.
For the first problem,  the initial condition is
$$
(\rho, u, p)=\cases{(1, -2, 0.1) & if $0<x<0.5$ \\
\cr (1, 2, 0.1) & if $1>x>0.5$ \\},
$$
$\Delta x = 2.5\times 10^{-3}$ and the final time is $t=0.1$. For the second problem, the initial condition is
$$
(\rho, u, p)=\cases{(1, 0, 4\times 10^{-13}) & if $0<x<2-0.5\Delta x, \quad 2 + 0.5\Delta x < x <4$ \\
\cr (1, 0, 2.56\times 10^{8}) & if $2-0.5\Delta x<x<2 + 0.5\Delta x$ \\},
$$
$\Delta x = 5\times 10^{-3}$ and the final time is $t=10^{-3}$.

\begin{figure}[p]
\begin{center}
\includegraphics[width=1.2\textwidth]{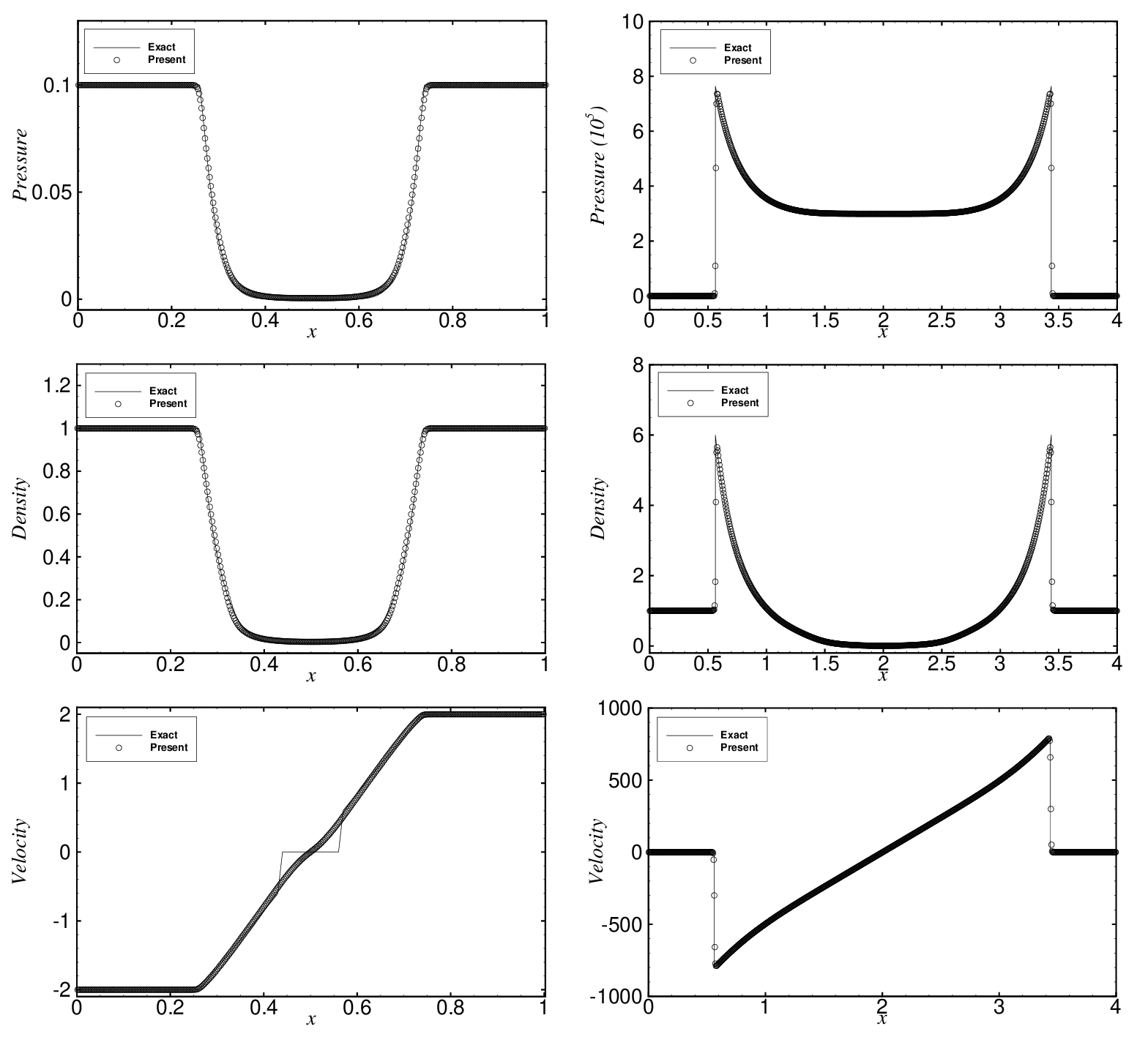}
\caption{One-dimensional problems involving vacuum or near vacuum:
(left) double rarefaction problem; (right) planar Sedov blast-wave problem.} \label{1d}
\end{center}
\end{figure}

Figure \ref{1d} gives the computed pressure, density and velocity distributions, which show good agreement with the exact solutions. %
Although a vacuum occurs in the solution of the double rarefaction problem, the results still exhibit accurate density and pressure profiles in the rarefaction-wave regions. As a vacuum occurs, the solution at the center of the domain strictly speaking has no physical meaning. Note that compared to Zhang and Shu \cite{zhang2011positivity} (see their Fig. 5.1 (right)) for the planar Sedov blast-wave problem a slightly sharper blast wave is obtained in the present results. This may be due to the fact that Zhang and Shu \cite{zhang2011positivity} have modified the original Lax-Friedrichs flux to use a single maximum signal speed other than the respective maximum eigenvalues. %
\subsection{Two-dimensional problems involving vacuum or near vacuum}
We consider two two-dimensional problems involving vacuum or near vacuum.
The first problem is the two-dimensional Sedov problem which has been studied in Zhang and Shu \cite{zhang2010positivity, zhang2011positivity}.
The computation is performed on the domain $[0,0]\times[1.1, 1.1]$, where a high pressure region occupies the computation cell at the lower-left corner.
The initial condition is given by
$$
(\rho, u, v, p)=\cases{(1, 0, 0, 4\times 10^{-13}) & if $x > \Delta x, \quad y > \Delta y$ \\
\cr (1, 0, 0, \frac{9.79264}{\Delta x \Delta y}\times 10^{4}) & else \\},
$$
where $\Delta x = \Delta y = 1.1/160$. The final time is $t = 1.0 \times 10^{-3}$. A reflective boundary condition is applied at the lower and left boundaries, and an outflow condition is applied at the right and upper boundaries. %
\begin{figure}[p]
\begin{center}
\includegraphics[width=1.2\textwidth]{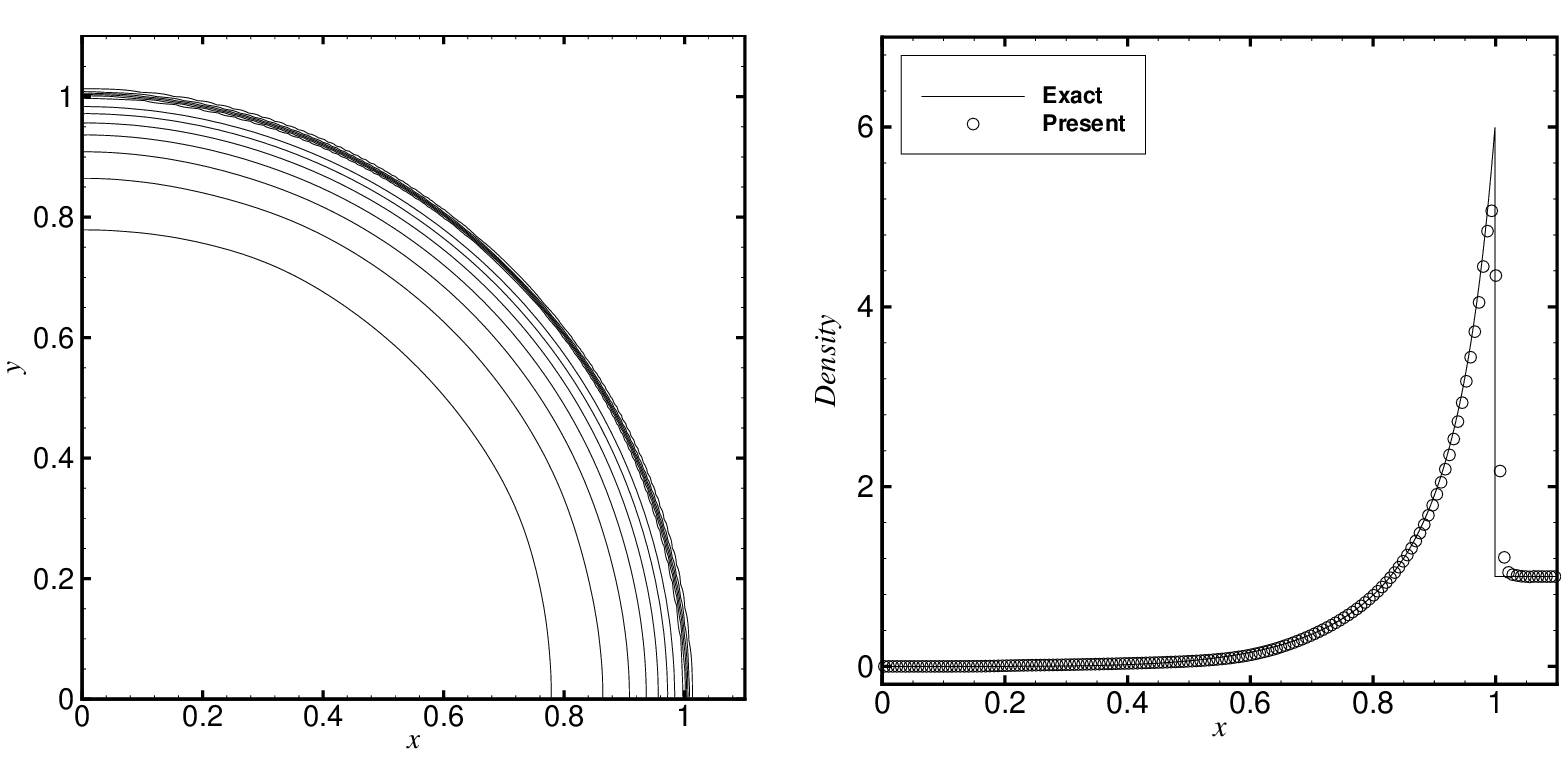}
\caption{Two-dimensional Sedov problem:
(left) 10 density contours from 0 to 6; (right) density profile along $y = 0$.} \label{2d-sedov}
\end{center}
\end{figure}
Figure \ref{2d-sedov} gives the computed density profiles. One can observe that these results are in very good agreement with the exact solution.

\begin{figure}[p]
\begin{center}
\includegraphics[width=1.2\textwidth]{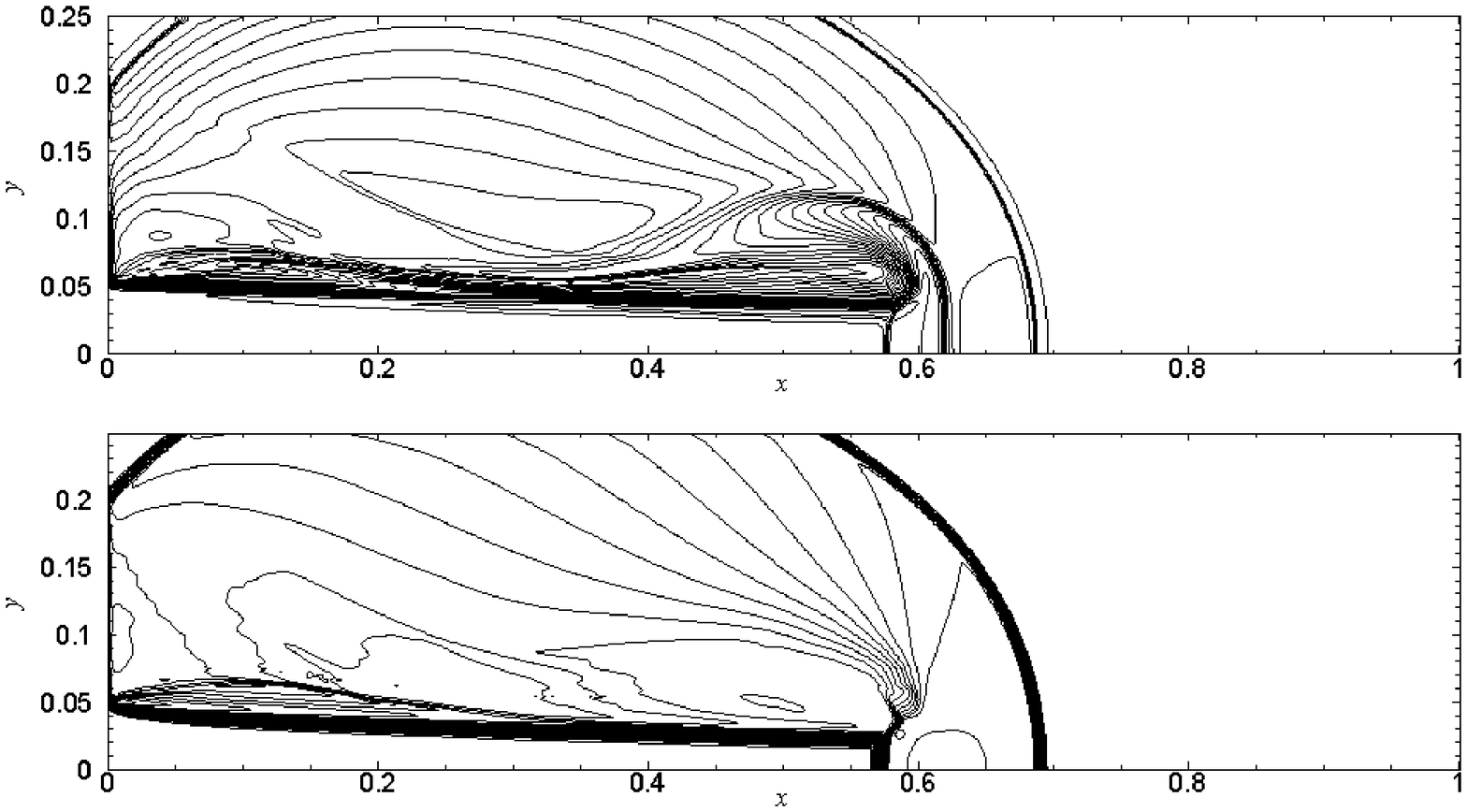}
\caption{Mach-2000 jet problem:
(upper) 30 density contours of logarithmic scale from -4 to 4; (lower) 30 pressure contours of logarithmic scale from -1 to 13.} \label{2d-mach-2000}
\end{center}
\end{figure}

The second problem is the Mach-2000 jet problem, which has been computed in Zhang and Shu \cite{zhang2010positivity, zhang2011maximum, zhang2011positivity}. The computation is performed on the domain $[0,1]\times[0, 0.25]$,
Initially, the entire domain is filled with ambient gas with $(\rho, u, v, p) = (0.5, 0, 0, 0.4127)$.
A reflective condition is applied at the lower boundary,
an outflow condition is applied at the right and upper boundaries, and an inflow condition is applied at the left boundary with states $(\rho, u, v, p) = (5, 800, 0, 0.4127)$ if $y<0.05$ and $(\rho, u, v, p) = (0.5, 0, 0, 0.4127)$ otherwise.
A CFL number of 0.25 is used and the final time is 0.001. Since $\gamma = 5/3$ is used, the speed of the jet 800 gives about Mach 2100 with respect to the sound speed in the jet gas. Figure \ref{2d-mach-2000} gives the computed density and pressure profiles in logarithmic scale. %
One can observe that these results are in very good agreement with those in
Zhang and Shu  \cite{zhang2010positivity} (their Fig. 4.6) computed with the same resolution. %
\subsection{One-dimensional problems involving very strong discontinuities}
We show that, combined with the proposed flux limiters, the WENO-CU6-M1 scheme passes two one-dimensional test problems, which cannot be computed with the original scheme without limiting, involving very strong discontinuities: the two blast-wave interaction problem \cite{woodward1984numerical}, and the Le Blanc problem \cite{loubcre2005subcell}. The latter is an extreme shock-tube problem.
For the first problem, the
initial condition is
$$
(\rho, u, p)=\cases{(1, 0, 1000) & if $0<x<0.1$ \\ \cr
(1, 0, 0.01) & if $0.1<x<0.9$ \\
\cr (1, 0, 100) & if $1>x>0.9$
\\},
$$
$\Delta x = 2.5\times 10^{-3}$, and the final time is $t=0.038$. Reflective boundary conditions are applied at both $x=0$ and
$x=1$. The reference ``exact" solution is a high-resolution 
numerical solution
on $3200$ grid points calculated by the WENO-CU6 
scheme \cite{hu2010adaptive}.
For the second problem, the initial condition is
$$
(\rho, u, p)=\cases{(1, 0, \frac{2}{3}\times 10^{-1}) & if $0<x<3$ \\
\cr (10^{-3}, 0, \frac{2}{3}\times 10^{-10}) & if $3<x<9$},
$$
$\gamma = 5/3$, $\Delta x = 9/800$ and the final time is $t=6$.

\begin{figure}[p]
\begin{center}
\includegraphics[width=1.2\textwidth]{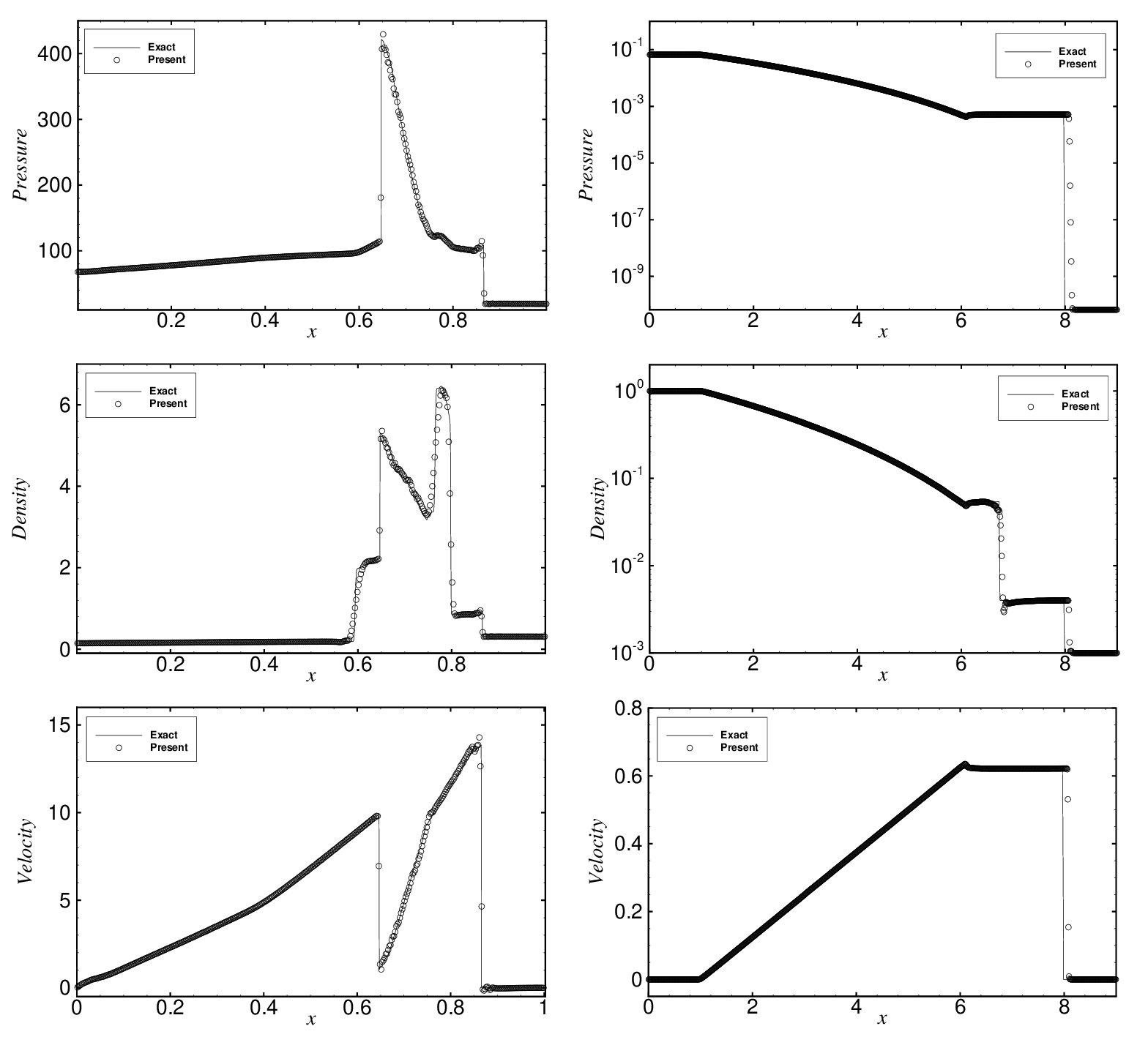}
\caption{One-dimensional problems involving very strong discontinuities:
(left) two blast wave problem; (right) Le Blanc shock-tube problem.} \label{1d-strong}
\end{center}
\end{figure}

Figure \ref{1d-strong} gives the
computed pressure, density and velocity distributions, although at relatively low resolution, which show a good agreement with the exact or
reference solutions. %
The magnitudes of the small over-shoots (see Fig. \ref{1d-strong}(left)) 
and the small errors at the shock position 
(see Fig. \ref{1d-strong}(right)) decrease when the resolution is
increased (not shown here). For the two blast-wave interaction 
problem the present results
are comparable to those obtained by the WENO-CU6-M2 scheme 
\cite{hu2011scale} at the same resolution (see their Fig. 3). Note that the WENO-CU6-M2 scheme stabilizes for very strong discontinuities in a different way, but still cannot compute the Le Blanc problem. %
\subsection{Two-dimensional problems involving very strong discontinuities}
We first consider the problem from Woodward and Colella \cite{woodward1984numerical} on the double Mach reflection of a strong shock.
A Mach 10 shock in air is reflected from the wall with incidence angle of $60^{\circ}$. The initial condition is
$$
(\rho, u, v, p)=\cases{(1.4, 0, 0, 1) & if $y < 1.732(x - 0.1667)$ \\
\cr (8, 7.145, -4.125, 116.8333) & else \\
},
$$
and the final time is $t = 0.2$. The computational domain for this problem is $[0,0]\times[4,1]$.
Initially, the shock extends from the point $x = 0.1667$ at the bottom
to the top of the computational domain.
Along the bottom boundary, at $y = 0$, from $x = 0$ to $x = 0.1667$
the post-shock conditions are imposed, whereas a reflective condition
is set from $x = 0.1667$ to $x = 4$. Inflow and outflow conditions
are applied at the left and right boundaries, respectively.
The states at the top boundary are set to describe the exact
motion of a Mach 10 shock.
\begin{figure}[p]
\begin{center}
\includegraphics[width=1.2\textwidth]{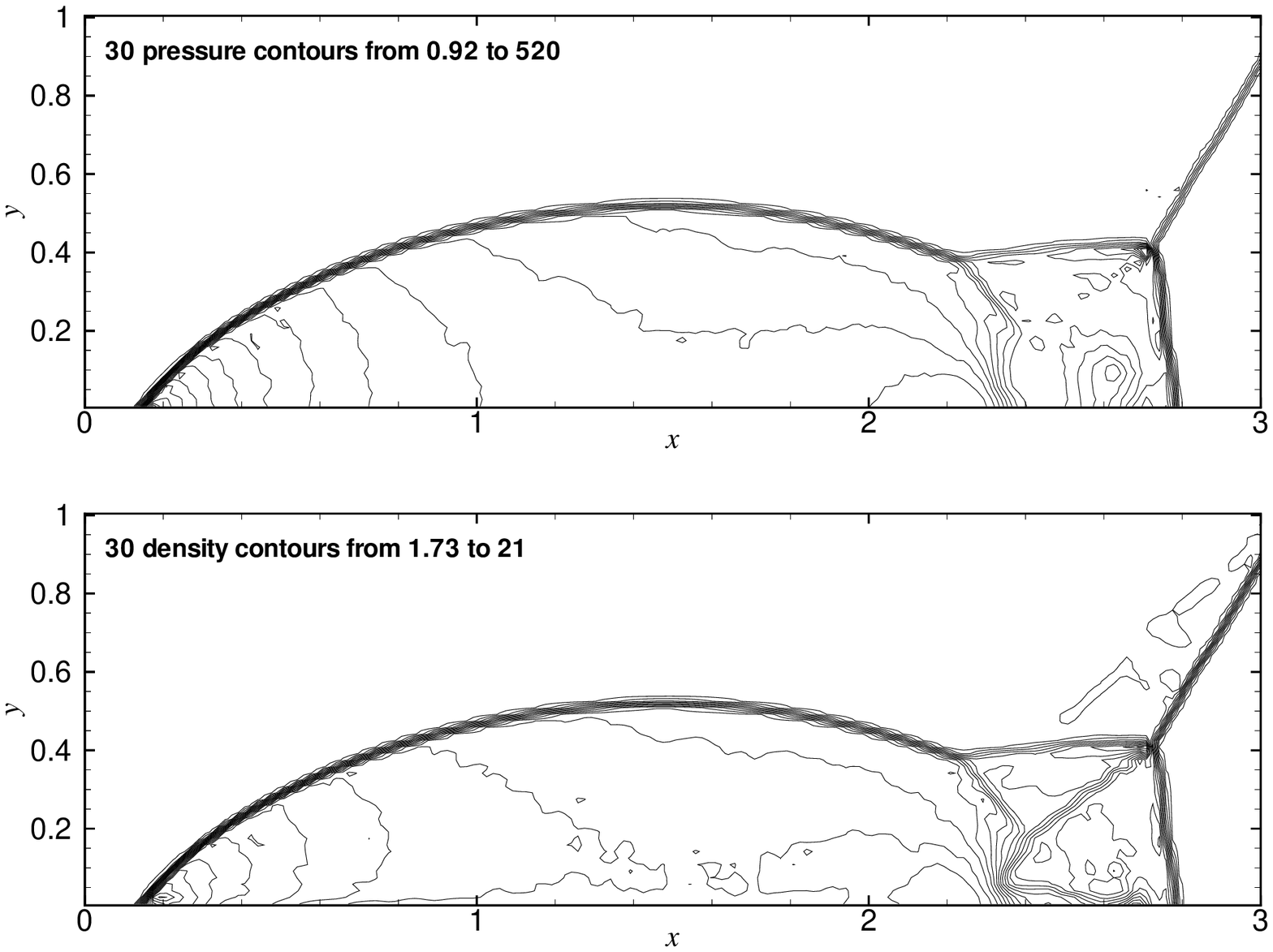}
\caption{Double-Mach reflection of a Mach 10 shock wave at $t = 0.2$:
(upper) 30 pressure contours from 0.92 to 520; (lower) 30 density contours from 1.73 to 21.} \label{double-mach}
\end{center}
\end{figure}
Figure \ref{double-mach} shows the pressure and density contours of the solution on a $240\times 60$ grid.
Note that compared to the results obtained by WENO-CU6-M2 \cite{hu2011scale}  (their Fig. 4) a good agreement is observed. Especially, both predict a strong near-wall jet, which is usually smeared in the previous computations with the same resolution \cite{kim2005high, kawai2008localized, hu2010adaptive}.

\begin{figure}[p]
\begin{center}
\includegraphics[width=0.75\textwidth]{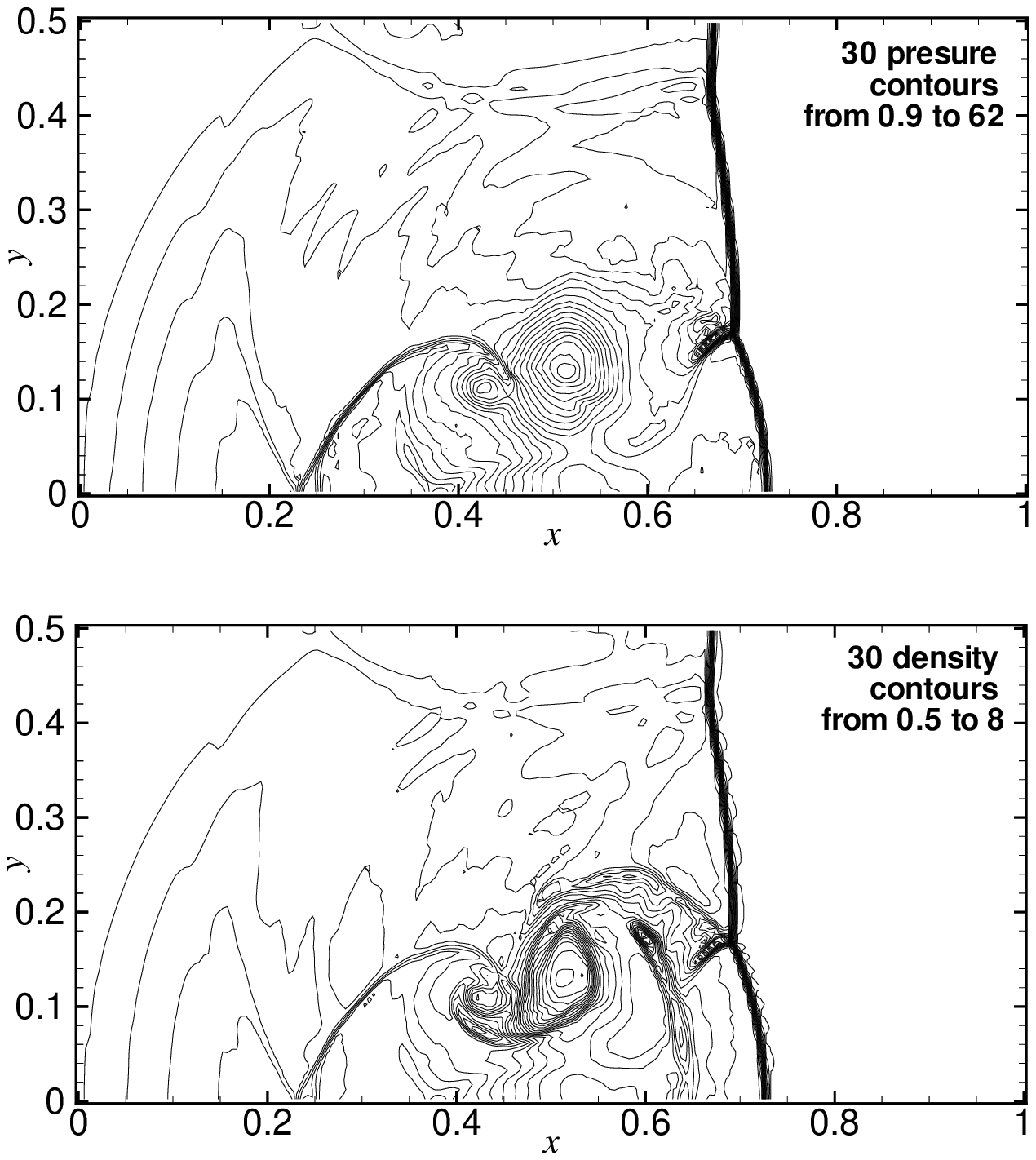}
\caption{Shock-bubble interaction problem at $t = 0.15$:
(left) 30 pressure from 0.9 to 62; (right) 30 density contours from 0.5 to 8.} \label{mach-6}
\end{center}
\end{figure}

We then consider a shock-bubble interaction problem,
when a Mach 6 shock wave in air impacts on a cylindrical helium bubble. Air and helium are treated as the same ideal gas fluid for simplicity.
Numerical computations for this problem can be found in Bagabir and Drikakis \cite{bagabir2001mach}. The initial conditions are
\begin{equation}
\cases{(\rho=1, u=-3, v=0, p=1) & pre-shocked air \\
\cr (\rho=5.268, u=2.752, v=0, p=41.83) & post-shocked air \\
\cr (\rho=0.138, u=-3, v=0, p=1) & helium bubble \\
},\label{eq:sgn}
\end{equation}
and the final time is $t = 0.15$. The computational domain for this problem is $[0,0]\times[1,0.5]$.
Initially, the shock wave is at $x=0.05$, and the half helium bubble of radius 0.15 is at (0,0.25).
Note that a frame velocity $u = -3$ is applied to keep the bubble approximately in the center of the computational domain.
Reflective conditions are applied at the lower and upper boundaries,
an outflow condition is applied at the right boundary, and an inflow condition is applied to the left boundary with the post-shocked state.
Figure \ref{mach-6} shows the pressure and density contours of the solution on a $200\times 100$ grid.
These results show a fairly good agreement with those in Bagabir
and Drikakis \cite{bagabir2001mach} (their Fig. 6) at the same resolution. The secondary reflected shock wave
and triple-wave configurations are calculated with good resolution. Note that since the WENO-CU6-M1 scheme has smaller numerical dissipation
than the MUSCL scheme used in Bagabir and Drikakis \cite{bagabir2001mach},
the present results show a less smeared bubble interface and more detailed structures near the triple-wave region. %
\section{Concluding remarks}
In this paper we have proposed a very simple method
to enforce the positivity-preserving property for general high-order conservative schemes.
The method first detects critical numerical fluxes which may lead to negative density and pressure,
then limits the fluxes to satisfy a sufficient condition for preserving positivity.
Though an extra time-step size 
condition is required to maintain the formal order of accuracy,
it is less restrictive than those in previous works.
In addition, since the method uses the general form of a conservative scheme, similarly as the approaches of Zhang and Shu \cite{zhang2011positivity}, it can be applied to flows with a general equation of state and source terms in a straightforward way.
\section*{Acknowledgment}
We thank Dr. Xiangxiong Zhang for inspirational discussions.
The research of the third author is supported by 
AFOSR grant F49550-12-1-0399 and NSF grant DMS-1112700.
\bibliographystyle{plain}

\begin{thebibliography}{10}

\bibitem{bagabir2001mach}
A.~Bagabir and D.~Drikakis.
\newblock Mach number effects on shock-bubble interaction.
\newblock {\em Shock Waves}, 11(3):209--218, 2001.

\bibitem{einfeldt1991godunov}
B.~Einfeldt, C.D.~Munz, P.L.~Roe, and B.~Sjogreen.
\newblock {On Godunov-type methods near low densities}.
\newblock {\em J. Comput. Phys.}, 92(2):273--295, 1991.

\bibitem{gressier1999positivity}
J.~Gressier, P.~Villedieu, and J.M. Moschetta.
\newblock Positivity of flux vector splitting schemes.
\newblock {\em J. Comput. Phys.}, 155(1):199--220, 1999.

\bibitem{hu2011scale}
X.Y.~Hu and N.A.~Adams.
\newblock Scale separation for implicit large eddy simulation.
\newblock {\em J. Comput. Phys.}, 2011.

\bibitem{hu2004kinetic}
X.Y.~Hu and B.C.~Khoo.
\newblock Kinetic energy fix for low internal energy flows.
\newblock {\em J. Comput. Phys.}, 193(1):243--259, 2004.

\bibitem{hu2010adaptive}
X.Y. Hu, Q.~Wang, and N.A. Adams.
\newblock {An adaptive central-upwind weighted essentially non-oscillatory
  scheme}.
\newblock {\em J. Comput. Phys.}, 229(23):8952--8965, 2010.

\bibitem{jiang1996cient}
G.-S. Jiang and C.-W. Shu.
\newblock {Efficient implementation of weighted ENO schemes}.
\newblock {\em J. Comput. Phys}, 126:202--228, 1996.

\bibitem{kawai2008localized}
S.~Kawai and S.K.~Lele.
\newblock Localized artificial diffusivity scheme for discontinuity capturing
  on curvilinear meshes.
\newblock {\em J. Comput. Phys.}, 227(22):9498--9526, 2008.

\bibitem{kim2005high}
D.~Kim and J.H. Kwon.
\newblock A high-order accurate hybrid scheme using a central flux scheme and a
  weno scheme for compressible flowfield analysis.
\newblock {\em J. Comput. Phys.}, 210(2):554--583, 2005.

\bibitem{loubcre2005subcell}
R.~Loub{\v{c}}re and M.J. Shashkov.
\newblock A subcell remapping method on staggered polygonal grids for
  arbitrary-lagrangian-eulerian methods.
\newblock {\em J. Comput. Phys.}, 209(1):105--138, 2005.

\bibitem{perthame1996positivity}
B.~Perthame and C.-W. Shu.
\newblock {On positivity preserving finite volume schemes for Euler equations}.
\newblock {\em Numerische Mathematik}, 73(1):119--130, 1996.

\bibitem{sedov1959similarity}
L.I.~Sedov.
\newblock {\em Similarity and Dimensional Methods in Mechanics (Similarity and
  Dimensional Methods in Mechanics, New York}.
\newblock Academic Press, 1959.

\bibitem{shu1988efficient}
C.-W. Shu and S.~Osher.
\newblock {Efficient implementation of essentially non-oscillatory
  shock-capturing schemes}.
\newblock {\em J. Comput. Phys.}, 77(2):439--471, 1988.

\bibitem{SO2}
C.-W. Shu and S.~Osher.
\newblock {Efficient implementation of essentially non-oscillatory
  shock-capturing schemes, II}.
\newblock {\em J. Comput. Phys.}, 83:32--78, 1989.

\bibitem{tao1999gas}
T.~Tang and K.~Xu.
\newblock {Gas-kinetic schemes for the compressible Euler equations:
  positivity-preserving analysis}.
\newblock {\em Zeitschrift f\"{u}r Angewandte Mathematik und Physik (ZAMP)},
  50(2):258--281, 1999.

\bibitem{toro2009riemann}
E.F. Toro.
\newblock {\em Riemann solvers and numerical methods for fluid dynamics: a
  practical introduction}.
\newblock Springer Verlag, 2009.

\bibitem{woodward1984numerical}
P.~Woodward and P.~Colella.
\newblock The numerical simulation of two-dimensional fluid flow with strong
  shocks.
\newblock {\em J. Comput. Phys.}, 54(1):115--173, 1984.

\bibitem{zhang2010positivity}
X.~Zhang and C.-W. Shu.
\newblock {On positivity preserving high order discontinuous Galerkin schemes
  for compressible Euler equations on rectangular meshes}.
\newblock {\em J. Comput. Phys.}, 229:8918--8934, 2010.

\bibitem{zhang2011maximum}
X.~Zhang and C.-W. Shu.
\newblock Maximum-principle-satisfying and positivity-preserving high-order
  schemes for conservation laws: survey and new developments.
\newblock {\em Proceedings of the Royal Society A: Mathematical, Physical and
  Engineering Science}, 467:2752--2776, 2011.

\bibitem{zhang2011positivity}
X.~Zhang and C.-W. Shu.
\newblock {Positivity-preserving high order finite difference WENO schemes for
  compressible Euler equations}.
\newblock {\em J. Comput. Phys.}, 231:2245--2258, 2012.

\end{thebibliography}

\end{document}